\documentclass[a4paper,fleqn,usenatbib]{mnras}

\usepackage[T1]{fontenc}
\usepackage{ae,aecompl}

\usepackage{graphicx}	% Including figure files
\usepackage{amsmath}	% Advanced maths commands
\usepackage{amssymb}	% Extra maths symbols
\usepackage{graphics}
\usepackage{graphicx}
\usepackage{epsfig}
\usepackage{txfonts}
\usepackage{blindtext}
\usepackage{longtable}

\newcommand{\ci}{\ion{C}{i}}	% per cm-squared

\title[Dust in GRB C\textsc{i} absorbers]{On the dust properties of high redshift molecular clouds and the connection to the 2175\,\AA~extinction bump}

\author[K.~E.~Heintz et al.]{K.~E.~Heintz$^{1}$,\thanks{E-mail: keh14@hi.is}
	T.~Zafar$^{2}$,
	A.~De~Cia$^{3}$,
	S.~D.~Vergani$^{4,5}$,
	P.~Jakobsson$^{1}$,
	J.~P.~U.~Fynbo$^{6,7}$,
	\newauthor
	D.~Watson$^{6,7}$,
	J.~Japelj$^{8}$,
	P.~M{\o}ller$^3$,
	S.~Covino$^{9}$, 
	L.~Kaper$^{8}$ \&
	A.~C.~Andersen$^{7}$ 
	\\
	$^{1}$Centre for Astrophysics and Cosmology, Science Institute, University of Iceland, Dunhagi 5, 107 Reykjav\'ik, Iceland \\
	$^{2}$Australian Astronomical Optics, Macquarie University, 105 Delhi Road, North Ryde, NSW 2113, Australia \\
	$^{3}$European Southern Observatory, Karl-Schwarzschildstrasse 2, D-85748 Garching bei M\"unchen, Germany \\
	$^4$GEPI, Observatoire de Paris, PSL Research University, CNRS, Place Jules Janssen, 92190 Meudon \\
	$^5$Institut d'Astrophysique de Paris, CNRS-UPMC, UMR7095, 98bis bd Arago, 75014 Paris, France \\
	$^6$Cosmic Dawn Center (DAWN) \\
	$^7$Niels Bohr Institute, University of Copenhagen, DK-2100 Copenhagen, Denmark \\
	$^8$Anton Pannekoek Institute for Astronomy, University of Amsterdam, Science Park 904, 1098 XH Amsterdam, The Netherlands \\
	$^9$INAF -- Osservatorio Astronomico di Brera, Via E. Bianchi 46, I-23807 Merate (LC), Italy 
}

\date{Accepted XXX. Received YYY; in original form ZZZ}

\pubyear{2019}

\begin{document}
\label{firstpage}
\pagerange{\pageref{firstpage}--\pageref{lastpage}}
\maketitle

\begin{abstract}
We present a study of the extinction and depletion-derived dust properties of gamma-ray burst (GRB) absorbers at $1<z<3$ showing the presence of neutral carbon (\ci). By modelling their parametric extinction laws, we discover a broad range of dust models characterizing the GRB {\ci} absorption systems. In addition to the already well-established correlation between the amount of {\ci} and visual extinction, $A_V$, we also observe a correlation with the total-to-selective reddening, $R_V$. All three quantities are also found to be connected to the presence and strength of the 2175\,\AA~dust extinction feature. 
While the amount of {\ci} is found to be correlated with the SED-derived dust properties, we do not find any evidence for a connection with the depletion-derived dust content as measured from [Zn/Fe] and $N$(Fe)$_{\rm dust}$. 
%Contrary to these SED-derived dust properties, the amount of {\ci} is not found to be correlated with the depletion-derived dust content, derived from [Zn/Fe] and $N$(Fe)$_{\rm dust}$. 
To reconcile this, we discuss a scenario where the observed extinction is dominated by the composition of dust particles confined in the molecular gas-phase of the ISM. We argue that since the depletion level trace non-carbonaceous dust in the ISM, the observed extinction in GRB {\ci} absorbers is primarily produced by carbon-rich dust in the molecular cloud and is therefore only observable in the extinction curves and not in the depletion patterns. This also indicates that the 2175\,\AA~dust extinction feature is caused by dust and molecules in the cold and molecular gas-phase. This scenario provides a possible resolution to the discrepancy between the depletion- and SED-derived amounts of dust in high-$z$ absorbers.
\end{abstract}


%%%%%%%%%%%%%%%%%%%%%%%%%%%%%%%%%%%%%%%%%%%%%%%%%%

%%%%%%%%%%%%%%%%% BODY OF PAPER %%%%%%%%%%%%%%%%%%

%\clearpage

\section{Introduction} \label{sec:intro}

One of the key constituents of galaxy formation and chemical evolution is the interplay and connection between dust, gas and metals in the interstellar medium (ISM). Since stars form in molecular clouds, identifying the cold neutral gas-phase in the ISM of galaxies in the early Universe will provide imperative clues to the onset of star formation. A powerful probe of the ISM in high-redshift galaxies are damped Lyman-$\alpha$ absorbers (DLAs), which can be observed toward background quasars \citep{Wolfe05} or in gamma-ray burst (GRB) host galaxies \citep{Jakobsson06,Fynbo09}.

Detailed information of the gas, metal and dust content of DLAs can be derived from absorption-line analyses or from the extinguished spectral energy distribution (SED). The optical afterglows of GRBs provide a unique tool to study the absolute extinction curves of their host galaxies because of their simple power-law spectra \citep[e.g.][]{Schady10,Greiner11,Zafar11,Covino13}, which is more difficult to constrain in quasar sightlines due to the more unknown variation in the underlying spectrum \citep[but see e.g.][]{Vladilo06,Jiang11,Ledoux15,Krogager16,Noterdaeme17,Fynbo17,Heintz18b}. In addition, the extinction curves provide information about the dust content, composition and grain size distribution compared to the attenuation of the GRB host \citep[which can be studied in emission after the afterglow has faded, e.g.][]{Corre18}, since the latter effect is a combination of reprocessed photons, dust-star geometry and radiative transfer effects \citep[][]{Narayanan18}. 

An alternative tracer of dust is the depletion of refractory elements such as silicon and iron from the gas-phase of the ISM \citep{Savaglio03,Jenkins09,DeCia16}. 
It is expected that the amount of dust derived from the SED, typically represented as the visual extinction, $A_V$, should then scale with the elements in the dust-phase \citep{Vladilo06,DeCia13}. However, there appears to be a discrepancy between the SED- and depletion-derived dust properties of GRB host absorption systems \citep{Savaglio04,DElia14,Wiseman17}. This inconsistency might be due to the fact that iron \citep[typically used to represent the dust-phase and argued to be the most reliable depletion measure;][]{DeCia18} does not trace most of the dust mass that will otherwise impact the inferred extinction \citep{Zafar13,Dwek16}. Another possibility is that the SED-derived visual extinction is underestimated if the typical grain size distribution is large, resulting in grey extinction \citep{Savaglio03}. We note though that \cite{Bolmer18} found that this tension is relieved at $z\gtrsim 4$, at least in H$_2$-bearing GRB absorbers, and that \cite{Zafar19} showed that $A_V$ is related to the total column density of refractory elements.

The cause of this putative discrepancy between the dust content inferred from the extinguished afterglow SED and the gas-phase depletion levels, might also be related to a too simplistic view of the location of the dust. Recent studies of GRB and quasar absorbers have for instance found that the visual extinction, $A_V$, and the strength of the 2175\,\AA\ extinction bump feature is correlated with the amount of neutral atomic-carbon \citep[\ion{C}{i};][]{Zafar12,Ledoux15,Ma18,Heintz19}. Absorption features from \ion{C}{i} are often found to be coincident with H$_2$ \citep{Srianand05,Jorgenson10,Noterdaeme18} and is therefore believed to be a good tracer of the cold and molecular gas-phase of the ISM. These relations indicate that \ion{C}{i} traces dust-rich systems and that the amount of cold, shielded gas and dust are connected. 

Carbonaceous material could produce the rare 2175\,\AA\ bump \citep{Stecher65,Weingartner01}, since carbonaceous grain growth requires neutral atomic-carbon and molecules in the ISM \citep{Henning98}. The 2175\,\AA~dust extinction feature is ubiquitously observed locally in the MW \citep{Fitzpatrick07} and in the LMC \citep{Gordon03}, but is mostly elusive in the SMC and in extragalactic sightlines \citep{Gordon03,Noll09,Zafar12,Zeimann15,Ledoux15,Ma18}. The strength and width of the characteristic feature vary with the surrounding environment but its central wavelength have been found to remain invariant \citep{Fitzpatrick86,Fitzpatrick07}. Early studies have suggested that graphite or polycyclic aromatic hydrocarbons (PAHs) are responsible for the 2175\,\AA~dust extinction feature \citep{Stecher65,Draine84,Leger89,Blanco96,Li97,Henning98,Weingartner01,Draine03}. More recently, \cite{Mishra17} reported that the strength of the 2175\,\AA~dust extinction feature is related to the amount of carbon in the dust-phase for a sample of 46 Galactic sightlines, further supporting carbonaceous dust grains as the carriers producing the 2175\,\AA~dust bump. 

In this paper, we study a sample of C\,\textsc{i} absorbers identified by \citet[][hereafter Paper I]{Heintz19} in GRB sightlines, with the aim of investigating the mineralogy, dust composition and grain size distribution of the cold and molecular gas-phase in these high-$z$ absorbers. Here, for the first time, we attempt to distinguish the different contributions of various elements that could be the carrier of the 2175\,\AA\ dust bump and affect the shape of the extinction curve at high redshifts. This sample of GRB {\ci} absorbers has the advantage that for example the total-to-selective reddening parameter, $R_V$, can be well constrained compared to similar systems in quasar sightlines. Spectroscopy of the optical afterglows of the GRBs in this sample (at a redshift range of $1 < z < 3$) thus provides a unique way to study the dust content of molecular clouds in star-forming galaxies at the peak of cosmic star formation. We have acquired elementel abundances and derived extinction curves for each of the GRB C\,\textsc{i} absorbers entering our sample, and combined with previous literature measurements we examine the distinct dust-phase of this unique sample of GRB C\,{\sc i} absorbers. 

\begin{table}
	\begin{minipage}{\columnwidth}
		\caption{GRB C\,\textsc{i} absorber sample. C\,\textsc{i}\,$\lambda\lambda$\,1560,1656 rest-frame equivalent widths are from Paper I.}
		\centering
		\begin{tabular}{lccc}
			\noalign{\smallskip} \hline \hline \noalign{\smallskip}
			GRB & $z_{\mathrm{GRB}}$ & $W_{\mathrm{r}}(\lambda\,1560)$ & $W_{\mathrm{r}}(\lambda\,1656)$ \\
			&& (\AA) & (\AA) \\
			\noalign {\smallskip} \hline \noalign{\smallskip}
			061121 & 1.3145 & $0.49\pm 0.04$ &  $0.25\pm 0.04$ \\
			070802 & 2.4511 & $0.70\pm 0.35$ & $1.36\pm 0.32$ \\
			080605 & 1.6403 & $0.64\pm 0.23$ &  $0.38\pm 0.15$ \\
			080607 & 3.0368 & $2.17\pm 0.08$ & $2.03\pm 0.04$ \\
			120119A & 1.7288 & $0.51\pm 0.05$ &  $0.77\pm 0.06$ \\
			120815A & 2.3581 & $0.12\pm 0.08$ & $0.21\pm 0.05$ \\
			121024A & 2.3024 & $0.08\pm 0.05$ & $0.11\pm 0.07$ \\
			150403A & 2.0571 & $0.34\pm 0.03$ & $0.50\pm 0.04$ \\
			180325A & 2.2486 & $0.58\pm0.05$ & $0.85\pm0.05$ \\
			\noalign{\smallskip} \hline \noalign{\smallskip}
		\end{tabular}
		%\tablefoot{}
		\label{tab:wr}
	\end{minipage}
\end{table}

We have structured the paper as follows. In Sect.~\ref{sec:sample}, we present the GRB C\,\textsc{i} absorber sample, including their basic individual properties such as redshift, gas-phase abundances and C\,\textsc{i} content. In Sect.~\ref{sec:met}, we describe our approach of computing the dust extinction laws and we provide our results in Sect.~\ref{sec:results}. We discuss a specific scenario on the location of the dust in the sightline to the GRB C\,\textsc{i} absorbers in Sect.~\ref{sec:disc}, and conclude on our work in Sect.~\ref{sec:conc}.
Throughout the paper, errors denote the $1\sigma$ confidence level. We assume a standard flat cosmology with $H_0 = 67.8$\,km\,s$^{-1}$\,Mpc$^{-1}$, $\Omega_m = 0.308$ and $\Omega_{\Lambda}=0.692$ \citep{Planck16}. Gas-phase abundances are expressed relative to the Solar abundance values from \cite{Asplund09}, where [X/Y] = $\log N(\mathrm{X})/N(\mathrm{Y}) - \log N(\mathrm{X})_{\odot}/N(\mathrm{Y})_{\odot}$, following the recommendations of \cite{Lodders09} on whether to use the photospheric or meteoritic values (or their average).

\begin{table*}
	\centering
	\begin{minipage}{0.9\textwidth}
		\centering
		\caption{Gas-phase abundances of the GRB C\,\textsc{i} absorbers. References for the individual bursts and gas-phase abundances are given in the Appendix under the dedicated notes on each GRB. \textbf{Notes.} *This value represents the lower limit on [O/H] measured from the equivalent width of O\,\textsc{i}\,$\lambda$\,1355, following \citet{Prochaska09}.}
		\begin{tabular}{lcccccccc}
			\noalign{\smallskip} \hline \hline \noalign{\smallskip}
			GRB & $\log N(\textsc{H\,i})$ & $\log N$(Zn\,\textsc{ii}) & $\log N$(Si\,\textsc{ii}) & $\log N$(Fe\,\textsc{ii}) & [Zn/H] & [Zn/Fe] & $\log N$(Fe)$_{\mathrm{dust}}$ & $\log N$(Si)$_{\mathrm{dust}}$ \\
			& (cm$^{-2}$) & (cm$^{-2}$) & (cm$^{-2}$) & (cm$^{-2}$) && & (cm$^{-2}$) & (cm$^{-2}$)  \\
			\noalign {\smallskip} \hline \noalign{\smallskip}
			061121 & $\cdots$ & $13.76\pm 0.06$ & $\cdots$ & $16.20\pm 0.03$ & $\cdots$ & $0.51\pm 0.07$ & $16.55\pm 0.06$ & $\cdots$ \\
			070802 & $21.50\pm 0.20$ & $13.60\pm 0.60$ &  $\cdots$ & $16.10\pm 0.10$ & $-0.46\pm 0.63$ & $0.45\pm 0.61$ & $16.36\pm 0.41$ & $\cdots$ \\
			080605 & $\cdots$ & $13.53\pm 0.08$ & $15.88\pm 0.10$ & $14.66\pm 0.11$ & $\cdots$ & $1.82\pm 0.14$ & $16.47\pm 0.08$ & $16.34\pm 0.09$ \\
			080607 & $22.70\pm 0.15$ & $>13.56$ &  $>16.21$ & $>16.48$ & $>-0.20$* & $\cdots$ & $\cdots$ & $\cdots$ \\
			120119A & $22.44\pm 0.12$ & $14.04\pm 0.25$ & $16.67\pm 0.35$ & $15.95\pm 0.25$ & $-0.96\pm 0.28$ & $1.04\pm 0.35$ & $16.95\pm 0.30$ & $16.69\pm 0.32$ \\
			120815A & $21.95\pm 0.10$ & $13.47\pm 0.06$ & $16.34\pm 0.16$ & $15.29\pm 0.05$ & $-1.04\pm 0.12$ & $1.13\pm 0.08$ & $16.39\pm 0.06$ & $15.58\pm 0.51$ \\
			121024A & $21.88\pm 0.10$ & $13.74\pm 0.03$ & $> 16.35$ & $15.82\pm 0.05$ & $-0.70\pm 0.10$ & $0.87\pm 0.06$ & $16.63\pm 0.03$ & $>16.44$ \\
			150403A & $21.73\pm 0.02$ & $13.32\pm 0.04$ & $>15.80$ & $15.54\pm 0.07$ & $-1.04\pm 0.04$ & $0.63\pm 0.08$ & $16.15\pm 0.05$ & $> 16.09$ \\
			180325A & $22.30\pm 0.14$ & $>14.09$ & $>16.12$ & $>16.68$ & $>-0.77$ & $\cdots$ & $\cdots$ & $\cdots$ \\
			\noalign{\smallskip} \hline \noalign{\smallskip}
		\end{tabular}
		\label{tab:metcol}
	\end{minipage}
\end{table*}

\section{Sample}\label{sec:sample}

The identification of GRB C\,\textsc{i} absorbers in Paper I was based on bursts observed as part of the GRB afterglow sample by \cite{Fynbo09} and the VLT/X-shooter GRB (XS-GRB) afterglow legacy survey \citep{Selsing18}. Neutral atomic-carbon was detected in ten GRB optical afterglow spectra obtained with low- to medium-resolution spectrograhps ($\mathcal{R} = 1000 - 10\,000$). The detection rate was found to be $\approx 25\%$ in the full statistical sample of GRBs compiled in Paper I at a completeness limit of $W_{\mathrm{r}}(\lambda\,1560) = 0.2$\,\AA. An overview of the GRBs with C\,\textsc{i} identified in absorption is provided in Table~\ref{tab:wr}, also including those bursts not entering the statistical sample from Paper I. The only C\,\textsc{i}-bearing GRB absorber from Paper I not included in this work is GRB\,060210. This burst was excluded from our analysis because the X-ray spectrum from the \textit{Swift}/XRT appears to be affected by X-ray flaring at the time when the optical spectrum was obtained. It is therefore not possible to constrain the intrinsic slope of the SED of this GRB, required to quantify the dust-extinction model. We note that \cite{Curran07} also find an offset between the X-ray and optical slopes based on their $R$- and $I$-band photometric data.

For all the nine GRB C\,\textsc{i} systems studied here, we provide the gas-phase abundances of H\,\textsc{i}, Zn\,\textsc{ii}, Si\,\textsc{ii}, and Fe\,\textsc{ii} in Table~\ref{tab:metcol}, where available from the literature. In the Appendix, an individual note on each GRB is given with references to where the literature values were obtained from. Based on the gas-phase abundances, we also list the derived metallicity, [Zn/H], the dust depletion, [Zn/Fe], and the dust-phase iron and silicon column density, $N$(Fe)$_{\mathrm{dust}}$ and $N$(Si)$_{\mathrm{dust}}$, for each GRB where possible. The actual metallicity of the C\,\textsc{i} systems are likely higher than reported here since Zn also depletes (although mildly) on dust grains. The dust-corrected metallicity can be computed as [M/H] = [Zn/H] + $\delta_{\mathrm{Zn}}$, where $\delta_{\mathrm{Zn}}$ is given by the level of depletion, assuming some empirically derived depletion sequence coefficients for Zn \citep[e.g.][]{DeCia16}. We will in the further analysis, however, only study [Zn/H] since it is a directly measured quantity, but note that the actual metallicity could be higher. Dust depletion patterns are gaining more prevalence as a universal measure of the dust content in the ISM of different environments (from the Milky Way to high-redshift DLAs) and specifically the relative abundance ratio of zinc to iron, [Zn/Fe], is found to be an ideal tracer of dust in the ISM \citep{DeCia18}. The dust-phase iron column density, $N$(Fe)$_{\mathrm{dust}}$, is in most cases also found to be correlated with the visual extinction, $A_V$, derived from the SED of the optical to X-ray afterglow spectra \citep{Vladilo06,DeCia13}, further supporting the use of depletion patterns as a tracer of dust in the ISM.

In Paper I, we found that C\,\textsc{i} was exclusively identified in GRB host absorption systems with large metal column densities ($\log N$(X) = $\log N$(H\,\textsc{i}) + [X/H] > 20.7), dust-phase iron column densities of $\log N$(Fe)$_{\mathrm{dust}} \gtrsim 16.2$ and constitutes the most dust-reddened population of GRBs at all redshifts. This is not unexpected, since a large metal and dust content is required to shield the neutral gas from the ambient galactic UV field. From Table~\ref{tab:metcol} we find that the neutral hydrogen column density ranges from $\log N$(H\,\textsc{i}) = 21.5 -- 22.7 and that the GRB C\,\textsc{i} absorbers have abundances between one-tenth of solar to solar values. We refer the reader to Paper I for a detailed discussion of the metal content of the GRB C\,\textsc{i} systems, where we also compared them to C\,\textsc{i}-bearing quasar absorbers. Here we focus on the SED-derived dust properties of the systems, which we derive for each individual burst (combined with literature measurements), and investigate how this relates to the depletion-derived dust properties in this type of cold gas absorbers. While this can in principle also be studied in quasar absorbers, it is typically much easier to constrain the specific dust-extinction models in sightlines to GRBs since their intrinsic spectra are simple, smooth power-laws.

%%%%%%%%%%%%%%%%%%%%%%%%%%%%%%%%%%%%%%%%%%%%%%%%%%%%%%%%%%%%%%%%%%%%%%%%%%%%
\section{SED analysis}    \label{sec:met}
%%%%%%%%%%%%%%%%%%%%%%%%%%%%%%%%%%%%%%%%%%%%%%%%%%%%%%%%%%%%%%%%%%%%%%%%%%% 

To model the parametric extinction laws of each of the individual GRBs in our sample, we follow the same procedure as \cite{Zafar18b}. We refer to their work for the explicit details of the fitting procedure, and only briefly summarize the methodology below. \citet{Zafar18b,Zafar18a} already determined the extinction curve parameters for the GRBs\,120119A, 120815A, 121024A, and 180325A entering our sample, so here we provide only the best fits for GRBs\,061121, 070802, 080605, 080607, and 150403A, using the same dust model for consistency.

The intrinsic GRB afterglow spectrum is believed to be emitted as synchrotron radiation from the interaction of the ultra-relativistic jet and the ISM. As a consequence, the intrinsic SED of the afterglow is expected to follow a power-law with $F_{\nu} = \nu^{-\beta}$. In some cases, a change in the spectral slope of $\Delta \beta = 0.5$ between the optical and X-ray spectra is observed \citep{Zafar11,Greiner11}, known as the cooling break \citep{Sari98}. The observed, dust extinguished spectrum can therefore be modeled as $F^{\mathrm{obs}}_{\nu} = F_{\nu}\,10^{-0.4\,A_{\lambda}}$, where $A_{\lambda}$ is the extinction due to dust absorption and scattering as a function of wavelength. 

\renewcommand{\arraystretch}{1.5}
\begin{table*}
	\centering
	\begin{minipage}{1.0\textwidth}
		\centering
		\caption{The best fit F\&M\,90 extinction curve parameters for the X-ray to the optical/near-infrared afterglow SEDs of the GRB C\,\textsc{i} absorbers. The best fit curve coefficients are from $^{a}$\citet{Zafar18b}, $^{b}$\citet{Zafar18a} or derived in this work. \textbf{Notes.} The strength of the 2175\,\AA~extinction bump is calculated as $A_{\mathrm{bump}} = \pi\,c_3 / (2\,\gamma\,R_V) \times A_V$. The reduced $\chi^2$, the number of degrees of freedom (dof), the null hypothesis probability (NHP) for the best fit, and the F-test probability is given in the Appendix for each individual GRB.}
		\begin{tabular}{lccccccccc}
			\noalign{\smallskip} \hline \hline \noalign{\smallskip}
			GRB & $c_1$ & $c_2$ & $c_3$ & $c_4$ & $\gamma$ & $x_0$ & $R_V$ & $A_V$ & $A_{\mathrm{bump}}$ \\
			&&&&&& ($\lambda^{-1}$) & & (mag) & (mag) \\
			\noalign {\smallskip} \hline \noalign{\smallskip}
			%060210* &&&&&&&&& \\
			061121 & $-4.63\pm 0.13$ & $2.11\pm 0.11$ & $<0.46$ & $0.34\pm 0.12$ & $1.00$ & $4.60$ & $2.88^{+0.23}_{-0.27}$ & $0.46^{+0.11}_{-0.15}$ & $<0.16$\\
			070802 & $0.03 \pm 0.23$ & $0.88\pm 0.17$ & $1.56\pm 0.27$ & $0.45\pm 0.09$ & $0.99\pm 0.01$ & $4.63\pm 0.01$ & $2.72^{+0.61}_{-0.54}$ & $1.20^{+0.15}_{-0.15}$ & $1.09\pm 0.34$ \\
			080605 & $-6.12\pm 0.84$ & $2.66\pm 0.39$ & $0.46\pm 0.11$ & $1.84\pm 0.50$ & $0.60\pm 0.08$ & $4.53\pm 0.01$ & $3.27^{+0.88}_{-0.92}$ & $0.48^{+0.13}_{-0.11}$ & $0.18\pm 0.08$ \\
			080607 & $1.14\pm 0.34$ & $0.36\pm 0.11$ & $3.05\pm 0.72$ & $0.17\pm 0.05$ & $1.22\pm 0.05$ & $4.51\pm 0.02$ & $4.14^{+1.05}_{-1.09}$ & $2.58^{+0.42}_{-0.45}$ & $2.45\pm 0.97$ \\
			120119A$^{a}$ & $-4.13\pm 0.08$ & $2.09\pm 0.07$ & $<0.43$ & $0.22\pm 0.10$ & $1.00$ & $4.60$ & $2.99^{+0.24}_{-0.18}$ & $1.02^{+0.11}_{-0.11}$  & $<0.26$\\
			120815A$^{a}$ &$-4.77\pm 0.08$ & $2.14\pm 0.07$ & $<0.32$ & $0.82\pm 0.08$ & $1.00$ & $4.60$ & $2.38^{+0.09}_{-0.09}$ & $0.19^{+0.04}_{-0.04}$ & $<0.05$\\
			121024A$^{a}$ & $-4.23\pm 0.06$ & $2.20\pm 0.08$ & $<0.35$ & $0.57\pm 0.05$ & $1.00$ & $4.60$ & $2.81^{+0.20}_{-0.16}$ & $0.26^{+0.07}_{-0.07}$ & $<0.06$ \\
			150403A & $-4.83\pm 0.08$ & $2.23\pm 0.05$ & $<0.35$ & $0.59\pm 0.02$ & $1.00$ & $4.60$ & $<2.81$ & $<0.13$ & $<0.03$\\
			180325A$^{b}$ & $-1.95\pm 0.39$ & $1.28\pm 0.17$ & $2.92\pm 0.19$ & $0.52\pm 0.19$ & $1.16\pm 0.06$ & $4.54\pm 0.03$ & $4.58^{+0.37}_{-0.39}$ & $1.58^{+0.10}_{-0.12}$ & $1.36\pm 0.19$ \\
			\noalign{\smallskip} \hline \noalign{\smallskip}
		\end{tabular}
		\label{tab:ext}
	\end{minipage}
\end{table*}

We used the spectral fitting package \texttt{XSPEC v. 12.9} to fit the rest-frame optical to near-infrared SEDs of the GRB afterglows, assuming a single or broken power-law together with the parametric dust model from \citet[][hereafter F\&M\,90]{Fitzpatrick90}. In systems where a broken power-law is preferred, a cooling break ($\nu_{\mathrm{break}}$) is required between the intrinsic spectral optical ($\beta_{\mathrm{opt}}$) and X-ray ($\beta_{\mathrm{X}}$) slopes and were fitted such that the change in slope is fixed to $\Delta \beta = 0.5$. The X-ray data are taken from the {\it Swift} archive facility \citep{Evans09} and for each case the spectrum in the 0.3 -- 10\, keV range is reduced around the SED mid-time with in {\tt XSELECT}. We then fix the total Galactic equivalent hydrogen column density, $N_{\mathrm{H,Gal}}$, for the X-ray spectra to the values from \cite{Willingale13} but leave the rest-frame GRB host galaxy equivalent hydrogen column density, $N_{\mathrm{H,X}}$, as a free parameter. The F\&M\,90 dust model allows the individual extinction curves to be fit through a set of eight parameters and is defined as
\begin{equation}
A_{\lambda} = \frac{A_V}{R_V}\,(c_1+c_2 x+c_3 D(x,x_0,\gamma)+c_4 F(x) + 1)~,
\end{equation}
where
\begin{equation}
D(x,x_0,\gamma) = \frac{x^2}{(x^2-x^2_0)^2 + x^2\gamma^2}
\end{equation}
and
\begin{equation}
F(x) = \begin{cases} 
0.539(x-5.9)^2 + 0.056(x-5.9)^3 &\mathrm{for}~x\ge 5.9  \\
0 &\mathrm{for}~x< 5.9
\end{cases}~,
\end{equation}
with $x=\lambda^{-1}$ in units of $\mu$m$^{-1}$. Basically, this dust model contains two components, one describing the linear UV part of the spectrum via the components $c_1$ (intercept), $c_2$ (slope) and the term $c_4F(x)$ describing the far-UV curvature. The second component is the Drude profile representing the 2175\,\AA~extinction bump, controlled by the parameters $c_3$ (bump strength), $x_0$ (central wavelength) and $\gamma$ (width of the bump). The last two parameters are the visual extinction, $A_V$, and the total-to-selective reddening, $R_V$. For the systems where the 2175\,\AA~extinction bump is not detected, we fix the Drude components to $\gamma = 1$\,$\mu$m$^{-1}$ and $x_0 = 4.6$\,$\mu$m$^{-1}$ but leave $c_3$ as a free parameter \citep[even though][found that fixing $c_3 = 0$ reduces the $\chi^2$]{Zafar15,Zafar18b}. We compute the $3\sigma$ upper limits on the bump strength instead, given by $c_3$, to compare to the GRBs where the 2175\,\AA~extinction bump is clearly detected. 

\begin{figure} %[!ht]
	\centering
	\epsfig{file=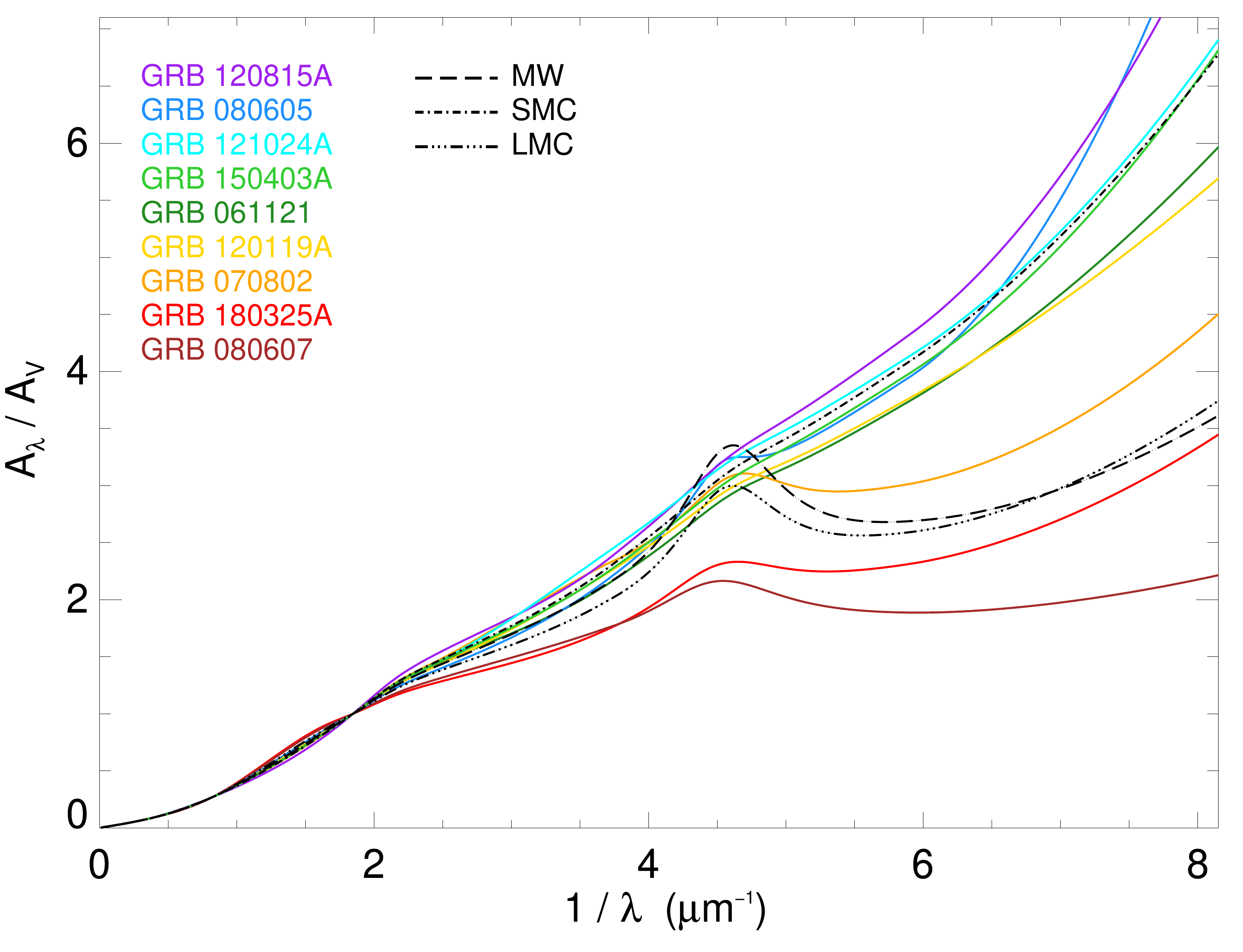,width=8.5cm}
	\caption{Individual extinction curves of the GRB C\,\textsc{i} absorbers. Each curve has been color-coded as a function of the total-to-selective extinction, $R_V$, violet representing a dust grain size distribution composed of small grains and red a dust composition with predominantly larger dust grain sizes. The 2175\,\AA~dust extinction feature appears to be more prominent for larger values of $R_V$. For comparison, the average extinction curves observed in the MW, SMC, and LMC are also shown.}
	\label{fig:ext}
\end{figure} 

%\newpage

%%%%%%%%%%%%%%%%%%%%%%%%%%%%%%%%%%%%%%%%%%%%%%%%%%%%%%%%%%%%%%%%%%%%%%%%%%%%
\section{Results}    \label{sec:results}
%%%%%%%%%%%%%%%%%%%%%%%%%%%%%%%%%%%%%%%%%%%%%%%%%%%%%%%%%%%%%%%%%%%%%%%%%%%

\begin{figure*} %[!ht]
	\centering
	\epsfig{file=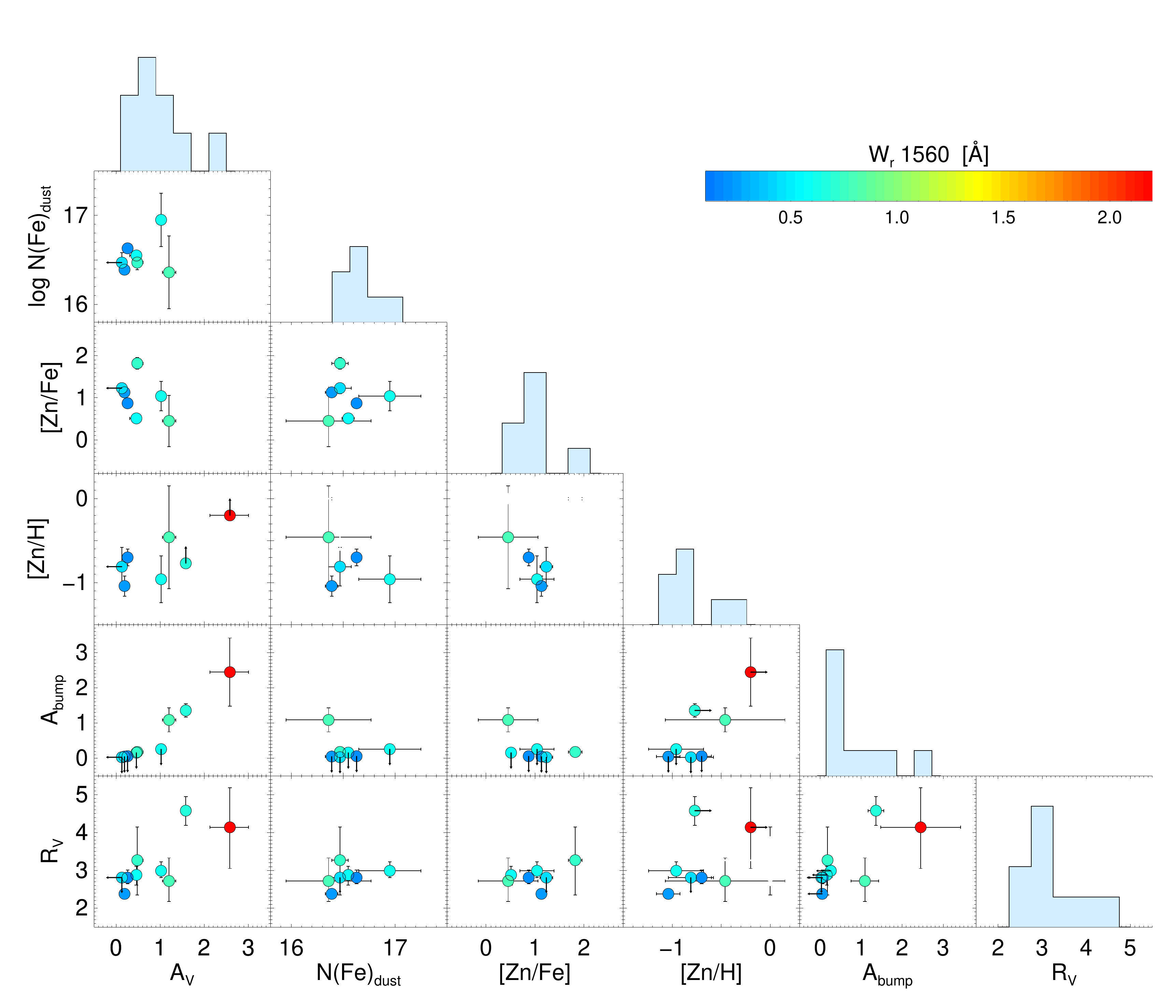,width=16cm}
	\caption{Pairwise scatter plots showing a comparison of different dust tracers, color-coded as a function of the rest-frame equivalent width of C\,{\sc i}\,$\lambda$\,1560. The SED-derived dust properties, i.e. $A_V$, $R_V$ and $A_{\rm bump}$ are compared to the commonly used depletion-derived dust tracers, the zinc-to-iron depletion, [Zn/Fe], and the dust-phase iron column density, $N$(Fe)$_{\rm dust}$. Positive trends with C\,{\sc i} are mainly observed for the SED-derived dust properties but also tentatively for the metallicity, [Zn/H] (see main text). For GRB\,080607 we show the estimated lower limit on the [O/H]-derived metallicity instead of [Zn/H], as also listed in Table~\ref{tab:metcol}.}
	\label{fig:dustcomp}
\end{figure*} 

Overall, we find a broad range of distinct dust models characterizing the GRB C\,{\sc i} absorbers. The individual best fit extinction curves are shown in Fig.~\ref{fig:ext}. In Table~\ref{tab:ext} we list the set of best fit parameters for the F\&M\,90 dust model for the GRBs 061121, 070802, 080605, 080607, and 150403A determined in this work, including previously published values for GRBs\,120119A, 120815A, 121024A, and 180325A from \citet{Zafar18b,Zafar18a}. We also provide $A_{\mathrm{bump}} = \pi\,c_3 / (2\,\gamma\,R_V) \times A_V$ (or the $3\sigma$ upper limit) for each of the GRB C\,\textsc{i} absorbers. The detection limit on the 2175\,\AA~dust extinction feature is typically $A_{\mathrm{bump}} < 0.2$\,mag. The equivalent hydrogen column density, $N_{\mathrm{H,X}}$, together with the reduced $\chi^2$, the number of degrees of freedom (dof), the null hypothesis probability (NHP) for the best fit, and the F-test probability (to determine preference over single or broken power-laws) is given in the Appendix for each of the GRBs examined in this work. The broken power-law model is considered a better fit for the cases where the F-test probability is smaller than 5\% \citep[see also][]{Zafar18b,Zafar18c}. 

In our sample, four out of nine cases ($\approx 45\%$) show a prominent 2175\,\AA~dust extinction feature (GRBs\,070802, 080605, 080607, and 180325A). Six systems ($\approx 65\%$) show relatively steep extinction curves with $R_V = 2.4 - 3.0$ (GRBs\,061121, 070802, 120119A, 120815A, 121024A, and 150403A), where the other three systems ($\approx 35\%$) show more "flat" extinction curves with $R_V = 3.3 - 4.5$ (i.e. GRBs\,080605, 080607, and 180325A). We also measure a large variation in bump strength between the individual systems, in the range $A_{\mathrm{bump}} = 0.2 - 2.5$\,mag.

In general, we find that the amount of C\,{\sc i} is connected to the dust properties inferred from the extinguished SED, such as $A_V$, $R_V$ and $A_{\rm bump}$. To quantify this we show pairwise scatter plots and histograms of different dust tracers color-coded as a function of $W_{\mathrm{r}}(\lambda\,1560)$ in Fig.~\ref{fig:dustcomp}. Due to the small sample size and relatively large error bars, the observed trends are only found to be correlated at $1 - 2\sigma$ significance. The connection of $A_V$ and $A_{\rm bump}$ with C\,{\sc i} is already well-established for quasar \citep{Ledoux15,Ma18} and GRB \citep{Zafar12,Heintz19} absorption systems, but here we include the additional information derived from the slope of the extinction, as inferred from $R_V$. The observed trend between $A_V$ and $W_{\mathrm{r}}(\lambda\,1560)$ indicates that larger amounts of cold gas is found in more dusty sightlines, where the additional trend with $R_V$ also suggests that the dust grain size distribution are on average larger in these sightlines as well. Contrary to this, we observe no relation between $A_V$, $R_V$ or the amount of {\ci} to the typically used depletion-derived dust tracers such as the zinc-to-iron depletion level, [Zn/Fe], and the dust-phase iron column density, $N$(Fe)$_{\mathrm{dust}}$ (see also Paper I).

We caution that since it is not possible to infer the depletion of the two strongest {\ci} absorbers (GRBs\,080607 and 180325A), the apparent non-correlation between the amount of {\ci} to [Zn/Fe] and $N$(Fe)$_{\mathrm{dust}}$ might be biased. However, even without those systems we are still probing a large range of depletion values (i.e. [Zn/Fe] = $0.5 - 2$), without seeing any trends with either $A_V$, $R_V$, $A_{\rm bump}$, or $W_{\mathrm{r}}(\lambda\,1560)$. Also, for the bursts where the depletion could be derived, the GRB showing the strongest extinction bump (GRB\,070802) shows the lowest value of [Zn/Fe] and $N$(Fe)$_{\rm dust}$. Finally, if we assume that e.g. GRB\,080607 has solar metallicity \citep{Prochaska09}, then the depletion of the system would be around [Zn/Fe] $\sim 1.1$ following the relation from \cite{Ledoux03}, which would still not produce any trend with $A_V$, $R_V$, $A_{\rm bump}$, or $W_{\mathrm{r}}(\lambda\,1560)$. We therefore argue that, even based on the small sample size, the amount of {\ci} and SED-derived dust properties is unlikely to be connected to the depletion-derived dust content for these particular absorbers.
In the following section we will attempt to provide a resolution to this apparent discrepancy.

\section{Discussion} \label{sec:disc}

\subsection{The presence of carbon-rich dust in molecular clouds}

The previously observed trend between $A_V$ and the amount of {\ci} indicates that whenever the GRB sightline intersects a molecular cloud, the amount of cold gas (C\,\textsc{i} in this case) is connected to the amount of dust derived from the extinction. Including the additional trend with $R_V$ observed in this work, it is clear that the average grain size distribution of the dust particles in the molecular cloud are also related to these properties. This would suggest that the C\,\textsc{i} systems are dust-rich with a composition dominated by carbonaceous material (as already speculated in Paper I), that the dust grain size distribution mainly consists of large dust grains, and that this carbon-rich dust dominates the shape of the extinguished SED. Here we argue for a simple scenario to explain these relations: The amount of C\,\textsc{i} simply reflects how deep into the intervening molecular cloud the GRB sightline probes, where we expect larger dust columns, dust grains and higher {\ci} column densities to be present closer to the centre of the cloud \citep{Bolatto13}. Below we will outline some of the key points that support this scenario. 

First, we argue that when cold and molecular gas is detected in GRB sightlines (i.e. from absorption features of C\,\textsc{i} and/or H$_2$) it is not related to the molecular gas associated with the GRB progenitor star. It has been found that the intense $\gamma$-ray flash, the afterglow emission or UV photons from the natal H\,\textsc{ii} region of the GRB will destroy dust and photo-dissociate H$_2$ (and therefore also ionize C\,\textsc{i}) out to $\approx10 - 100$\,pc \citep{Waxman00,Fruchter01,Draine02,Perna03,Whalen08}. In addition, \cite{Ledoux09} showed that photo-dissociation is effective out to distances of $\approx 500$\,pc from the GRB explosion site in the case of GRB\,050730. Since the physical sizes of dense, molecular gas clouds are often found to be small \citep[$l = 0.1 - 1$\,pc, e.g.][]{Balashev11,Krogager16}, any neutral and molecular gas in the vicinity of the GRB is likely completely ionized or photo-dissociated by the burst itself. Some molecular gas could be detected from the far end of giant molecular clouds though, if these are intersecting typical GRB sightlines.
Moreover, the bulk of the absorbing material in GRB hosts is typically found to be located at distances $0.5 - 2$\,kpc \citep{Vreeswijk07,Ledoux09,DElia09a,DElia09b} from the explosion site. For instance, \cite{Prochaska09} inferred at minimum distance of $d \gtrsim 100$\,pc of the molecular cloud observed towards GRB\,080607 based on the presence of Mg\,{\sc i} and {\ci}. Any intervening molecular cloud does therefore likely not belong to the molecular gas associated with the GRB progenitor star but is located further away (0.5 -- 2\,kpc) in the line-of-sight to the burst. \cite{Bolmer18} reached a similar conclusion and argued that this scenario could also explain why the strongest H$_2$-bearing GRB and quasar absorbers are more dust-depleted due to increased dust production and shielding in these systems, compared to the general observed trend between dust depletion and metal column density.

\begin{figure} %[!t]
	\centering
	\epsfig{file=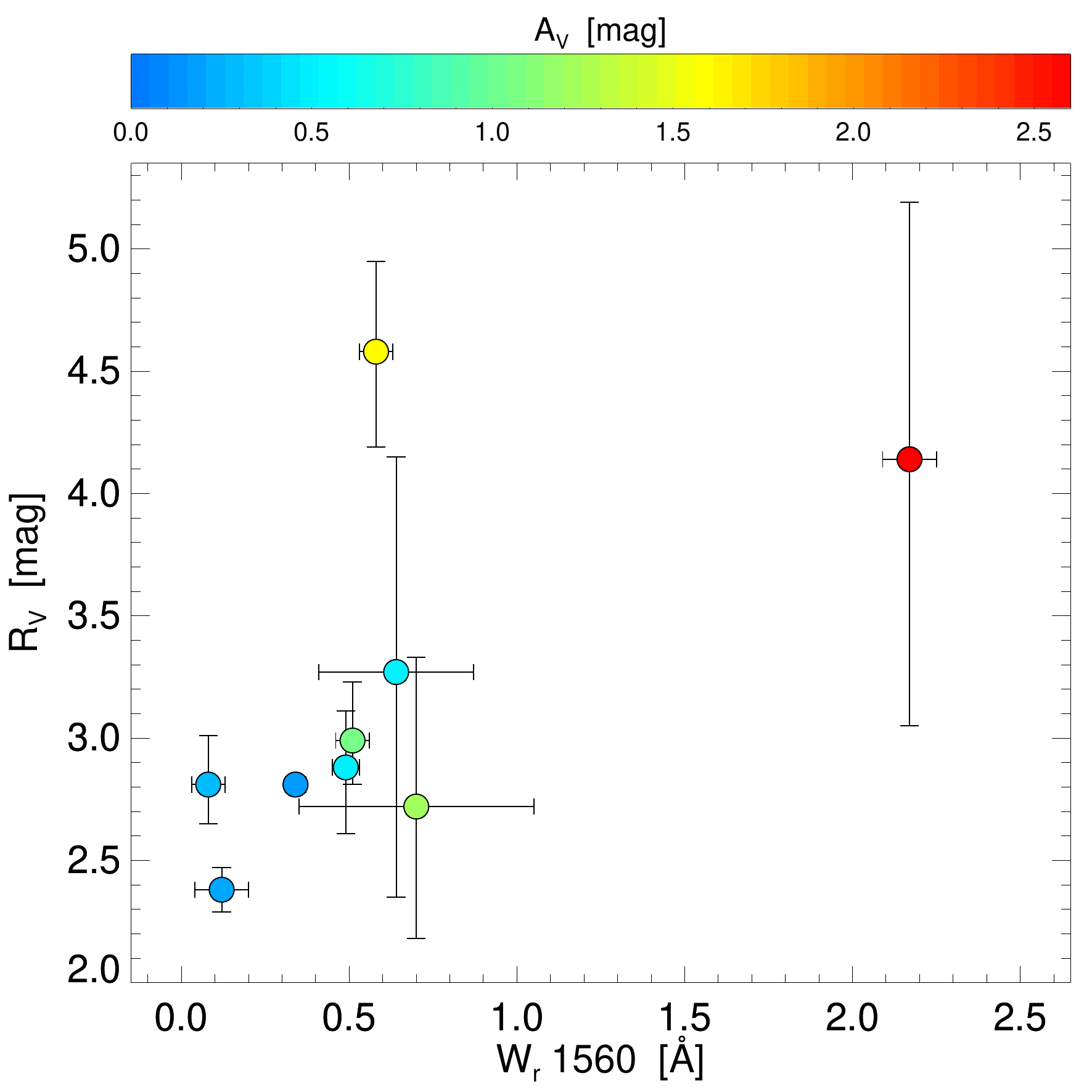,width=8.7cm}
	\caption{The total-to-selective extinction, $R_V$, as a function of the rest-frame equivalent width of C\,{\sc i}\,$\lambda$\,1560. The filled circles represent the GRB C\,\textsc{i} absorbers from Paper I studied in this work, color-coded as a function of the visual extinction, $A_V$. Increasingly dense environments expected for large amounts of C\,{\sc i} appears to be sensitive to the variation in the grain-size distribution and the total amount of dust.}
	\label{fig:wravrv}
\end{figure}   

\begin{figure*} %[!ht]
	\centering
	\epsfig{file=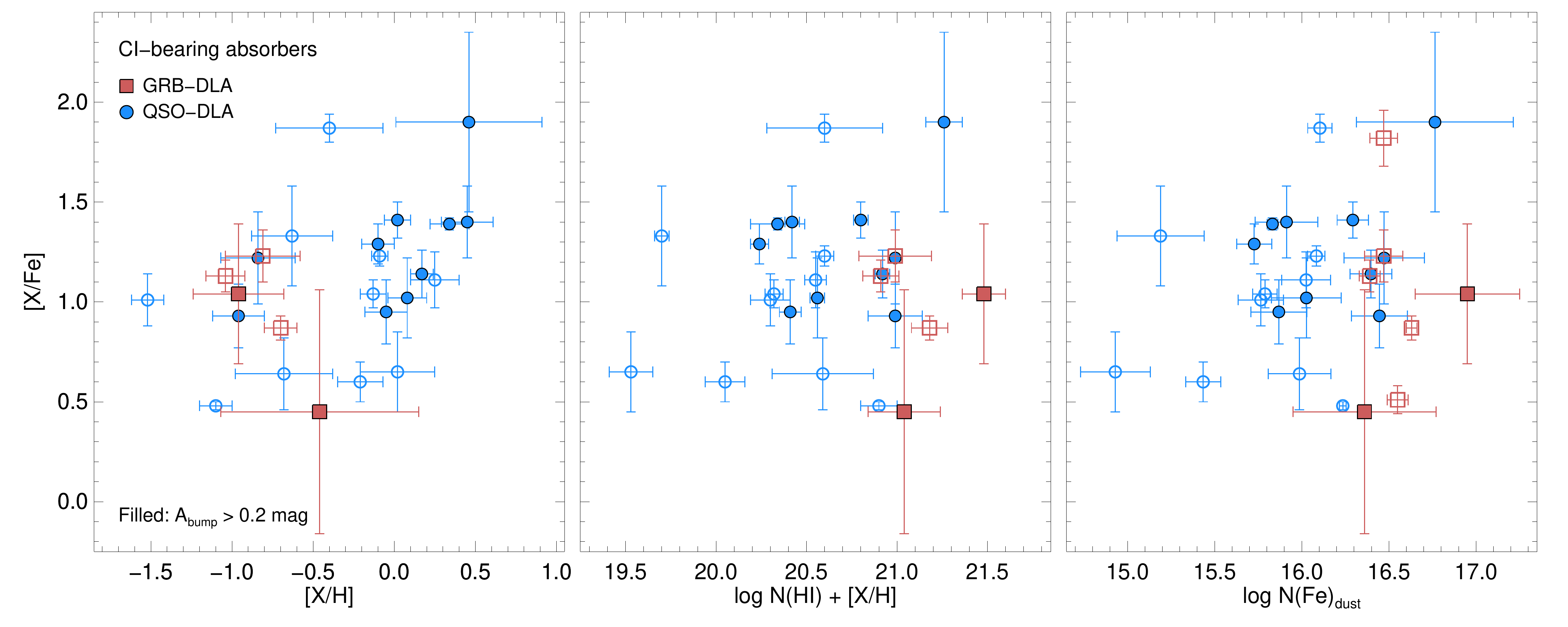,width=17cm}
	\caption{Dust depletion as a function of metallicity (left), metal column density (middle) and dust-phase iron column density (right) of quasar (blue) and GRB (red) C\,{\sc i} absorption systems. Filled symbols denote C\,{\sc i} absorbers where the 2175\,\AA~dust extinction feature has been securily detected with $A_{\rm bump} > 0.2$\,mag. There appears to be no correlation with the detection of the 2175\,\AA~dust bump to the gas and depletion-derived dust content of the absorbers. }
	\label{fig:2175dust}
\end{figure*} 

In this scenario, $A_V$ will as a consequence primarily trace dust located in the molecular gas-phase of the intervening absorber. Alternatively,
more massive galaxies with on average larger dust columns along the sightline through the ISM \citep{Kruhler11,Perley13,Zafar18c,Corre18}, could contain larger amounts of C\,{\sc i} due to more efficient shielding throughout the ISM. However, a massive galaxy with an associated dust-rich ISM would not necessarily contain larger dust grains (i.e. larger $R_V$). Indeed, no such correlation is observed in the Small and Large Magellanic Clouds (SMC and LMC), the Milky Way \citep[MW;][]{Gordon03,Fitzpatrick07} or in average high-redshift GRB absorbers \citep[excluding bursts with C\,\textsc{i} and/or H$_2$ detections;][]{Zafar18b,Zafar18c}. Since $A_V$ and $R_V$ appears to be correlated in our sample of GRB C\,\textsc{i} absorbers, it is a reasonable assumption that they are regulated by the same physical processes and thus originates from the same molecular cloud. This further supports that the SED-derived dust properties in GRB C\,{\sc i} absorbers characterize the dust composition and grain size distribution of the dust in the molecular gas-phase and not in ISM of the GRB host galaxies. 

The amount of C\,{\sc i} is then expected to correlate with the SED-derived dust extinction, but not necessarily with other dust tracers such as the dust depletion, [Zn/Fe], and the dust-phase iron column density, $N$(Fe)$_{\mathrm{dust}}$ (as observed, see Fig.~\ref{fig:dustcomp}). 
Since dust can form if there are sufficient dense and cold gas available, non-carbonaceous dust grains will specifically form via grain-growth in the ISM if there is a large reservior of metals available as well \citep[mostly O, Fe, Si, and Mg;][]{Draine03,DeCia16}, whereas carbon-rich dust will form if there is instead sufficient carbon available. Based on this, we therefore argue that in the ISM, the amount of dust is well-represented by either the derived depletion level, e.g. [Zn/Fe] \citep{DeCia18} and/or the dust-phase iron abundance, since these quantities trace non-carbonaceous dust. If this scenario is true, we predict that $A_V$ is only correlated to the non-carbonaceous dust tracers \citep[such as $N$(Fe)$_{\mathrm{dust}}$;][]{Vladilo06,DeCia13} if the GRB and quasar sightline does not intersect a molecular cloud. If instead, a molecular cloud consisting of significant amounts of carbon (in the form of C\,\textsc{i} or CO) intervenes the GRB and quasar sightline, the carbon-rich dust will significantly change the shape of the extinction curve, such that $N$(Fe)$_{\mathrm{dust}}$ or the depletion-derived $A_V$ is no longer a good representation of the integrated dust composition but only contributes mildly to the overall shape of the extinguished SED. This could also explain why \cite{Bolmer18} found that there is more carbon-rich dust in GRB hosts at $z > 4$, simply due to the smaller amounts of dust grains formed from grain growth in the ISM at this epoch.

Finally, we note that \cite{RamirezTannus18} found a correlation between the strength of diffuse interstellar bands \citep[DIBs; e.g.][]{Cami14} with the amount of dust towards the star-forming region M17, but an anti-correlation with $R_V$. The strongest DIBs are, however, believed to originate in the diffuse ISM, such that the anti-correlation with $R_V$ is likely only connected to the \lq warm\rq~gas-phase of the ISM and not to the molecular gas. We will discuss this further in Sect.~\ref{ssec:2175bump}.

\subsection{Dust composition of molecular clouds at high-$z$}

Following the arguments laid out above, it is now possible to constrain the dust composition and grain size distribution of these intervening, high-$z$ molecular clouds. First, we note from Fig.~\ref{fig:wravrv} that the variation in the grain-size distribution observed in this work support a model where small dust grains condense into large grains in increasingly dense environments \citep{Draine90,Weingartner01}. The relatively steep extinction curves observed toward GRBs\,061121, 070802, 120119A, 120815A, 121024A, and 150403A indicate a grain size distribution composed mainly of small ($\lesssim 0.25\,\mu$m) silicate grains in the ISM \citep{Mishra17}. The extinction curves observed toward GRBs\,080605, 080607, and 180325A, however, require large average grain sizes ($\approx 0.4-0.5\,\mu$m) and a significant contribution from small carbonaceoues dust grains to reproduce the observed 2175\,\AA~dust extinction feature (see the section below). 

We expect $R_V$ to be larger in these dust-rich GRB environments, simply because the intervening dust could have a large distribution of grain sizes and better conditions for growth in the more shielded inner regions of the molecular clouds. On the other hand, for sightlines probing the outer regions of the molecular clouds such a grain size distribution is also expected, since small grains are more easily destroyed by the ambient UV flux and therefore only the largest grains remain. This will in turn help to reduce the amount of ionizing photons that will photo-ionize {\ci} to C\,{\sc ii}.

\subsection{The origin of the 2175\,\AA~dust extinction feature} \label{ssec:2175bump}

Based on a sample of quasar absorbers selected solely on the basis of the 2175\,\AA~dust extinction feature, \cite{Ma18} found that all such systems are associated with strong absorption features from C\,{\sc i}. However, only $\approx 30\%$ of C\,{\sc i}-selected absorbers show the characteristic dust bump at a detectable level \citep{Ledoux15}. To further explore the conditions required for high-redshift, C\,{\sc i}-bearing absorbers to simultanously show the presence of the 2175\,\AA~dust extinction feature and whether this dust bump is produced in the ISM, we have compiled a sample of quasar C\,{\sc i} systems for which constraints of the extinction bump have been derived and compare them to the GRB C\,{\sc i} absorbers studied here, in terms of [Zn/Fe], [Zn/H], $\log N$(H\,{\sc i}) + [Zn/H], and $N$(Fe)$_{\rm dust}$, in Fig.~\ref{fig:2175dust}. The sample of C\,{\sc i}-bearing quasar absorbers is comprised of the C\,{\sc i}-selected systems from \cite{Ledoux15} with gas-phase abundances derived by \cite{Zou18}, and additional individual systems from the literature \citep[from][and Geier et al. subm.]{Guimaraes12,Krogager16,Krogager18,Noterdaeme17,Balashev17,Ranjan18}. While the C\,{\sc i} absorbers showing a significant 2175\,\AA~dust bump \citep[at more than $2\sigma$, i.e. $A_{\rm bump} > 0.2$\,mag;][]{Ledoux15} generally have large metal and depletion-derived dust contents, there appears to be no clear correlation for the detection probability of this feature with any of these absorption-line measured abundances. We note that \cite{Ma17} found a tentative anti-correlation between the bump strength and depletion (though mainly scatter-dominated), which is however not supported by the GRB and quasar C\,{\sc i} absorbers studied here. This suggests that the carriers producing the 2175\,\AA~dust extinction feature is also not located in the \lq warm\rq~neutral gas-phase of the ISM.

\begin{figure} %[!t]
	\centering
	\epsfig{file=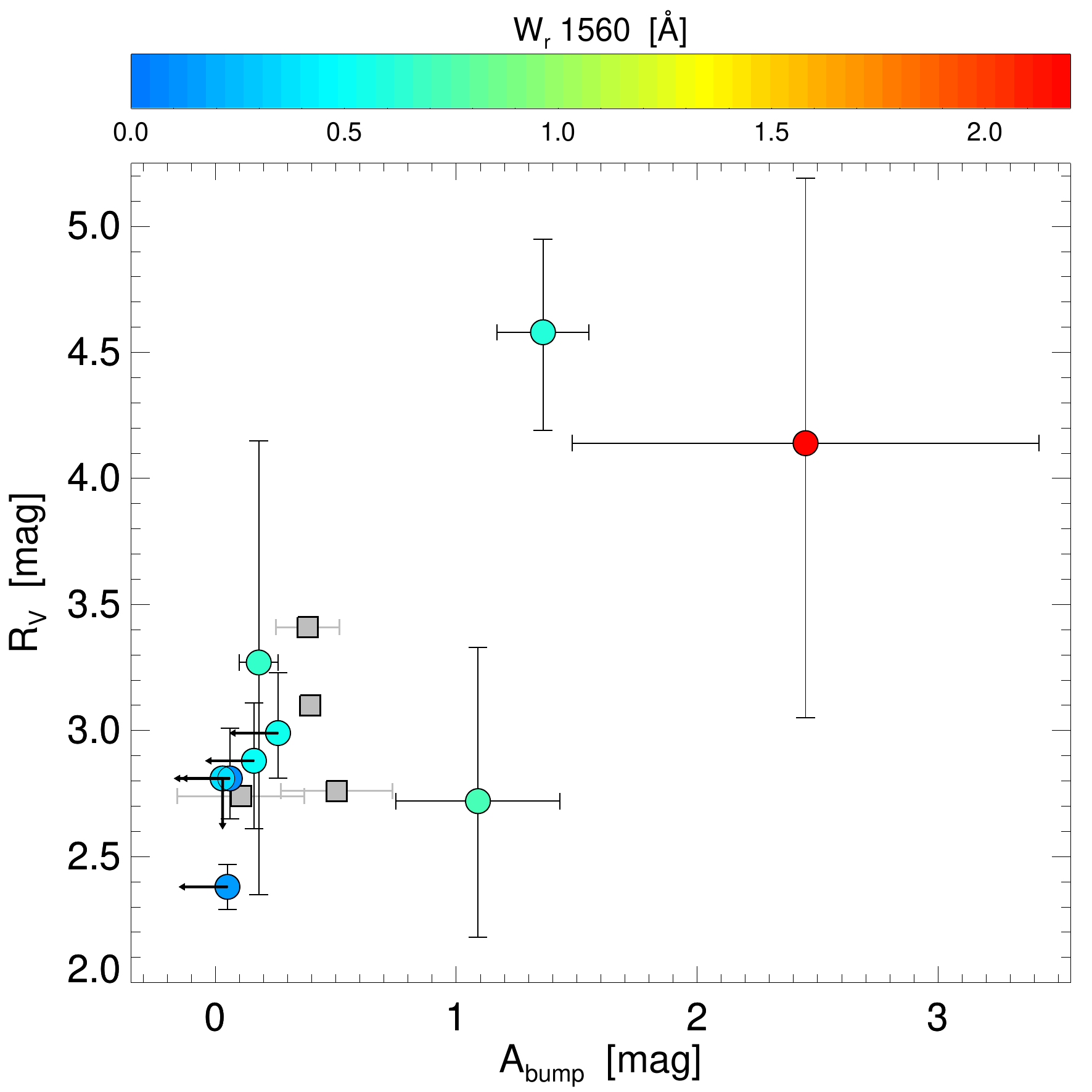,width=8.7cm}
	\caption{The total-to-selective extinction, $R_V$, as a function of the 2175\,\AA~bump strength, $A_{\rm bump}$. The colored filled circles represent the GRB C\,\textsc{i} absorbers from Paper I studied in this work, color-coded as a function rest-frame equivalent width of C\,{\sc i}\,$\lambda$\,1560. Grey squares show the mean bump strength of the quasar C\,{\sc i} absorbers from \citet{Ledoux15}, divided into the four different extinction curves used in the fit (represented by $R_V$) where the error bars denote the scatter in absorbers. The carriers producing the 2175\,\AA~dust extinction feature appears to be more prominent in systems with larger amounts C\,{\sc i} and with a grain size distribution composed of larger dust grains.}
	\label{fig:abrvav}
\end{figure}

If the 2175\,\AA~dust extinction feature is instead connected to the molecular gas-phase of the ISM, we would expect from the scenario outlined in the previous sections that the detection probability and strength of this feature is better correlated with the SED-derived dust properties.
In GRB and quasar sightlines, the detection and strength of the 2175\,\AA~dust extinction feature has previously been found to be connected to dusty sightlines \citep{Ledoux15,Ma18,Corre18} and the amount of C\,{\sc i} in the absorbing systems \citep{Zafar12,Ledoux15,Ma18,Heintz19}.  
In Fig.~\ref{fig:abrvav}, we explore the connection between the characteristic bump and the grain size distribution, represented by $R_V$ \citep{Draine03}. We find that the carriers producing the 2175\,\AA~dust extinction feature in GRB hosts appears to be more prominent in systems with large amounts of C\,{\sc i} and a grain size distribution composed mainly of large dust grains. The quasar C\,{\sc i} absorber population appears to also be consistent with this, though only limited information is available on the slope of the extinction, $R_V$ \citep{Ledoux15}. If $R_V$ and {\ci} is indeed connected to large dust grains and cold neutral gas observed deep into molecular clouds, the 2175\,\AA~dust extinction feature must by association be produced in the same molecular cloud.

Finally, \cite{Cortzen18} recently found that PAHs are connected to the molecular gas-phase in star-forming galaxies. If PAHs then contribute to or are the sole carrier of the 2175\,\AA~dust extinction feature \citep[as argued by][]{Li01,Draine07}, this would further support the scenario where the bump is produced by carriers located in molecular clouds. In addition, since the bump strength is not connected to the presence of DIBs \citep{Xiang11}, this excludes the scenario where the carriers of the 2175\,\AA~dust extinction feature is located in the \lq warm\rq~neutral ISM. This would also explain why GRB\,070802, showing a strong 2175\,\AA~extinction bump, stands out as a clear outlier in the relation between $A_V$ and $N$(Fe)$_{\rm dust}$ \citep{DeCia13}, simply because a significant dust component from the carbon-rich dust in the molecular cloud affects the shape of the SED but not the gas-phase depletions.

\section{Conclusions} \label{sec:conc}

In this work, we studied the dust properties of a sample of nine C\,{\sc i}-selected GRB absorption systems at $1<z<3$ with the goal of characterizing the mineralogy, dust composition and grain size distribution of the cold and molecular gas-phase in the ISM of these absorbers. This is the first time that a distinction between the various elements contributing to the observed extinction curve, in particular the 2175\,\AA~dust bump, has been attempted at high redshift. We derived parametric extinction curves for a subset of the sample, and combined with literature measurements we identified a broad range of dust models representing the full set of GRB C\,{\sc i} absorbers. In addition to the already established relations between the amount of C\,{\sc i} and the visual extinction, $A_V$, and the strength of the 2175\,\AA~extinction bump, $A_{\rm bump}$ \citep{Zafar12,Ledoux15,Ma18,Heintz19} we found evidence for an additional trend between the strength of the C\,{\sc i} absorption feature and the total-to-selective reddening parameter, $R_V$.
Contrary to this, we found no connection between the amount of C\,{\sc i} to the dust depletion, [Zn/Fe], or the dust-phase iron abundance, $N$(Fe)$_{\rm dust}$, but note a tentative correlation with the gas-phase metallicity, [Zn/H]. We, therefore, further advocate against using the depletion-derived dust content as a tracer of the total line-of-sight extinction \citep[as also previously cautioned;][]{Savaglio04,Wiseman17}, at least when the individual dust components cannot be securily identifed. 

We discussed a scenario where the main dust component causing the observed line-of-sight extinction in the GRB C\,{\sc i} absorbers is located in the intervening, C\,{\sc i}-bearing molecular cloud. We argued that such a scenario could explain the connection of C\,{\sc i} with only the SED-derived dust properties, and reconcile the apparent discrepancy between the extinction and depletion-derived dust properties. Here, the main component causing the observed extinction is carbon-rich dust in the molecular cloud which is not sensitive to the amount of depleted, non-carbonaceous metals in the dust-phase of the ISM. Consistently, we found that the detection and strength of the 2175\,\AA~dust extinction feature is also not linked to any of the non-carbonaceous dust indicators, but better correlated with $A_V$, $R_V$ and the amount of C\,{\sc i}. Moreover, the characteristic dust bump appears to be possibly linked to carbonaceous dust grains and the molecular gas-phase of the ISM, supporting PAHs as potential carriers of the 2175\,\AA~dust extinction feature. We also note that \citet{Mishra17} did not find any correlation between the extinction parameters and silicon in the dust-phase, but did find that the 2175\,\AA~dust bump scales with the amount of carbon in the dust-phase in Galactic sightlines. This is consistent with the above interpretation of the GRB absorbers showing significant {\ci} abundances.

The evidence for the validity of such a scenario in high-redshift {\ci}-bearing absorbers as presented here, is still only based on tentative correlations and should be verified if more GRB C\,{\sc i} systems, especially with the 2175\,\AA~dust extinction feature, are observed. 
Modelling the individual contributions from a large dust component located in the molecular gas-phase of the ISM and a secondary component in the \lq warm\rq~neutral medium would also be beneficial to further understand the observed dust content of high-redshift star-forming galaxies.

\section*{Acknowledgements}

We would like to thank the referee for providing a clear, constructive and concise report that greatly improved the interpretations presented here. We would also like to thank Karl Gordon for an enlightening discussion in the early stages of this work. KEH and PJ acknowledge support by a Project Grant (162948--051) from The Icelandic Research Fund. The Cosmic Dawn Center is funded by the DNRF.

%%%%%%%%%%%%%%%%%%%%%%%%%%%%%%%%%%%%%%%%%%%%%%%%%%



\bibliographystyle{mnras}
\bibliography{ref} % if your bibtex file is called example.bib

%%%%%%%%%%%%%%%%%%%%%%%%%%%%%%%%%%%%%%%%%%%%%%%%%%

%%%%%%%%%%%%%%%%% APPENDICES %%%%%%%%%%%%%%%%%%%%%

\appendix

\section{Notes on individual GRBs}

\subsection*{GRB\,061121}

\begin{figure}%[!ht]
	\centering
	\epsfig{file=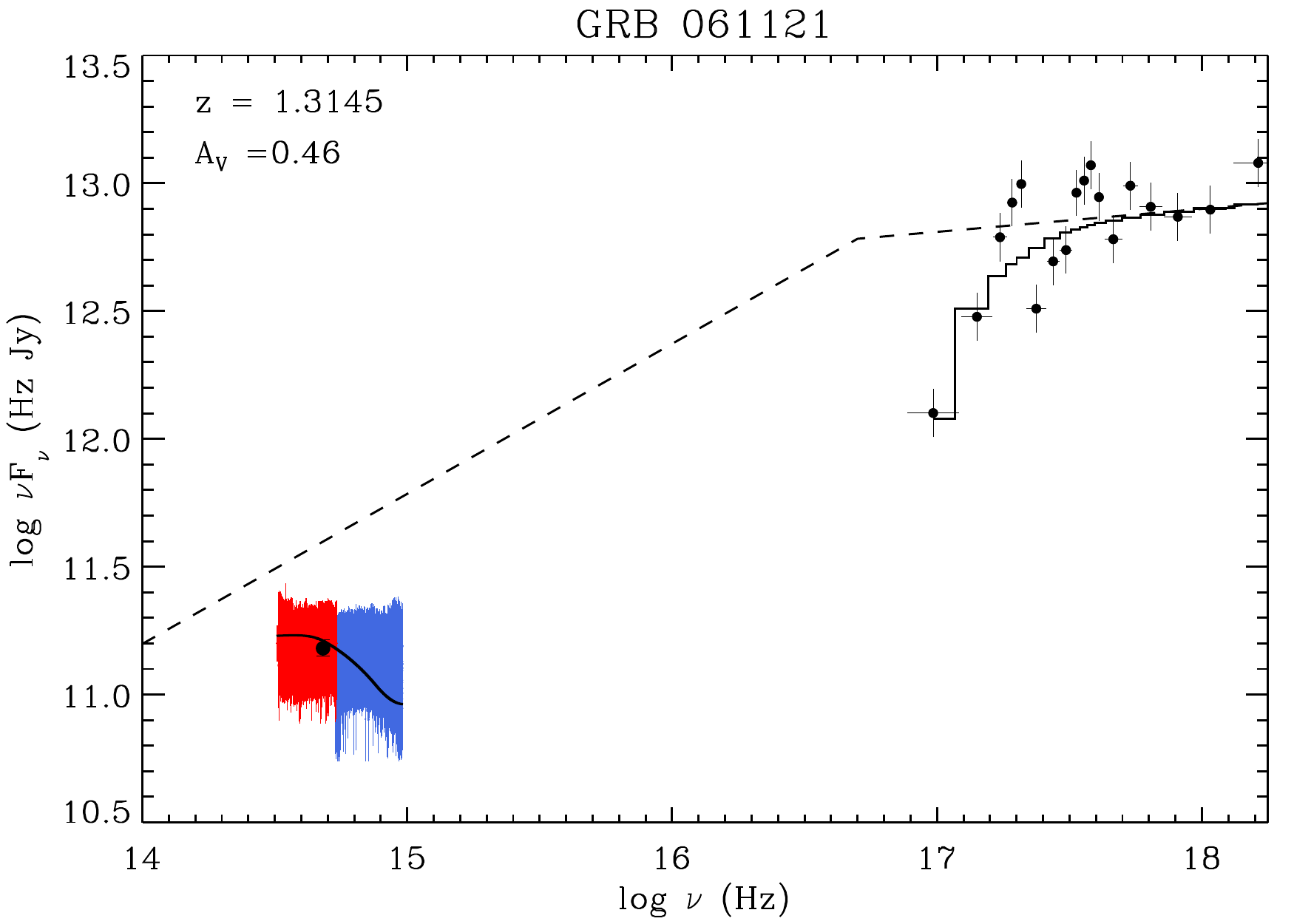,width=8.5cm}
	\caption{The observed afterglow SED of GRB\,061121. The \textit{Swift}/XRT data is indicated by the black points on the right. The blue and red colours correspond to the blue and red arm spectra obtained with the Keck/LRIS, respectively. The black overlaid point on the optical afterglow spectrum is the photometric point obtained from the acquisition camera. The errors on the spectroscopic and photometric data are also plotted. The best fit extinguished (solid lines) and intrinsic spectral slope (dashed line) are shown in black.}
	\label{afig:061121}
\end{figure}

This GRB was first detected by \textit{Swift} and reported by \cite{Page06}. For this case we derived the extinction curve parameters listed in Table~\ref{tab:ext} and the best fit model and $A_V$ are shown in Fig.~\ref{afig:061121}. The spectra examined in this work are obtained with the Keck/LRIS spectrograph and published by \cite{Fynbo09}. The gas-phase abundances listed in Table~\ref{tab:metcol} are from \cite{Zafar19}. In addition to the parametric dust model we also obtain the best fit for the equivalent neutral hydrogen column density of $N_{\mathrm{H,X}} < 0.94\times 10^{22}$\,cm$^{-2}$, the optical slope, $\beta_{\mathrm{opt}} = 0.41\pm 0.10$, the X-ray slope, $\beta_{\mathrm{X}} = 0.91\pm 0.11$, and the break frequency, $\log \nu_{\mathrm{break}} = 16.70\pm 0.15$\,Hz. The SED fits well with a broken power-law (with an F-test probability of <0.01) and the resulting reduced $\chi^2$ together with the number of degrees of freedom is $\chi^2/(\mathrm{dof}) = 1657/6161$, with a null hypothesis probability of $100\%$ for the best fit model. These results are consistent with the best fit models derived by \cite{Schady10} and \cite{Covino13}, which are based solely on photometry.

\subsection*{GRB\,070802}

\begin{figure} %[!ht]
	\centering
	\epsfig{file=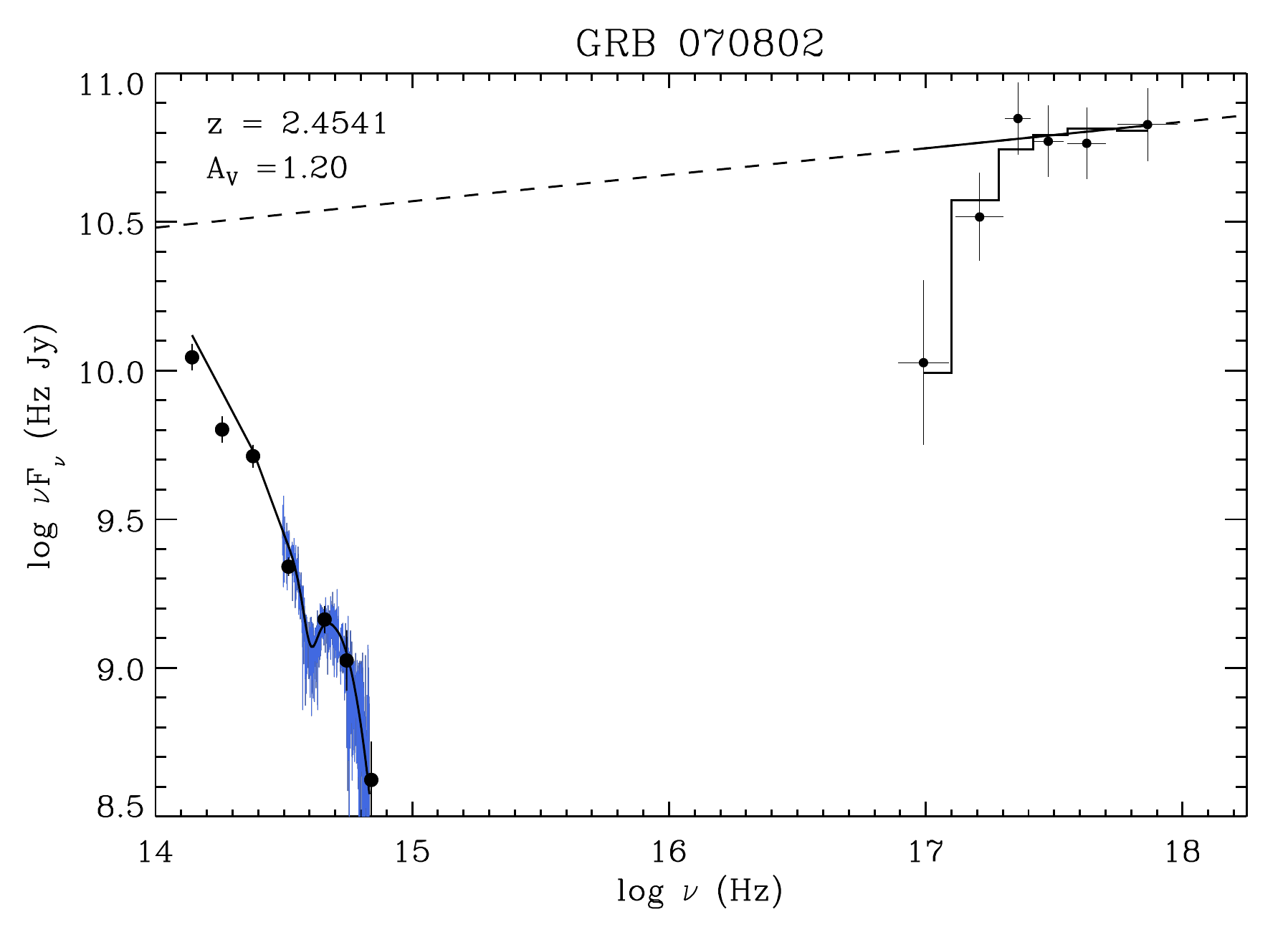,width=8.5cm}
	\caption{Same plot connotation as in Fig.~\ref{afig:061121} but for GRB\,070802 observed with the VLT/FORS2.}
	\label{afig:070802}
\end{figure}

This GRB was first detected by \textit{Swift} and reported by \cite{Barthelmy07}. For this case we derived the extinction curve parameters listed in Table~\ref{tab:ext} and the best fit model and $A_V$ are shown in Fig.~\ref{afig:070802}. The spectrum examined in this work is from the VLT/FORS2 spectrograph and published by \cite{Eliasdottir09,Fynbo09}. In addition to the parametric dust model we also obtain the best fit for the equivalent neutral hydrogen column density of $N_{\mathrm{H,X}} < 2.91 \times 10^{22}$\,cm$^{-2}$. The SED fits well with a single power-law (with an F-test probability of 1.00) with a slope of $\beta_{\mathrm{opt,X}} = 0.88\pm 0.06$ and the resulting reduced $\chi^2$ together with the number of degrees of freedom is $\chi^2/(\mathrm{dof}) = 750/1382$, with a null hypothesis probability of $100\%$ for the best fit model. This GRB shows a clear presence of the 2175\,{\AA} extinction bump \citep[as first reported by][]{Kruhler08,Eliasdottir09} and \cite{Zafar11} have fitted the parametric dust model prescribed by \cite{Fitzpatrick07} to the same optical afterglow examined here. Our results are consistent with their best fit model, but we will use the results from fitting the SED with the F\&M\,90 parametrization in this work for consistency.

\subsection*{GRB\,080605}

\begin{figure} %[!ht]
	\centering
	\epsfig{file=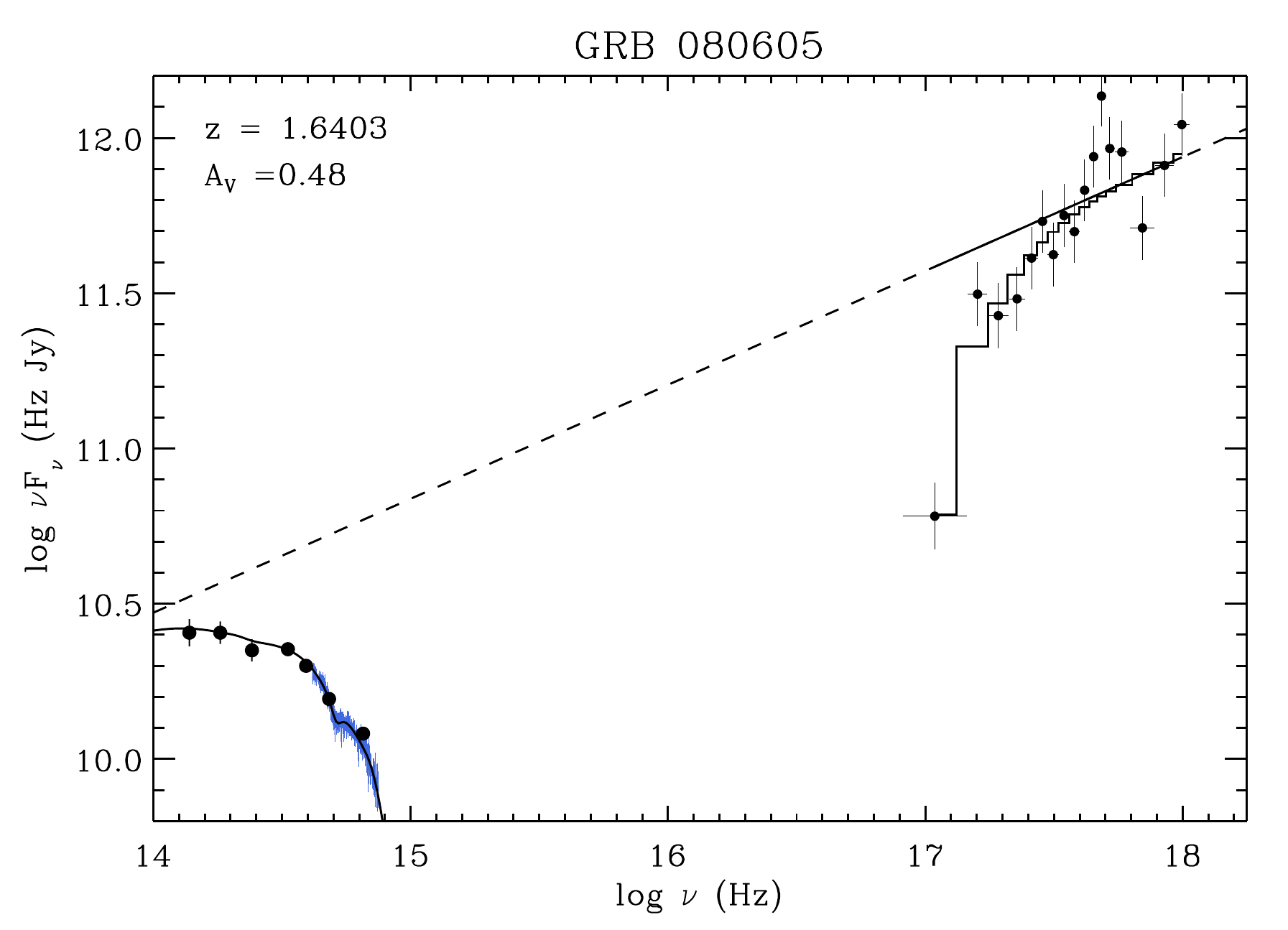,width=8.5cm}
	\caption{Same plot connotation as in Fig.~\ref{afig:061121} but for GRB\,080605 observed with the VLT/FORS2.}
	\label{afig:080605}
\end{figure}

This GRB was first detected by \textit{Swift} and reported by \cite{Sbarufatti08}. For this case we derived the extinction curve parameters listed in Table~\ref{tab:ext} and the best fit model and $A_V$ are shown in Fig.~\ref{afig:080605}. The spectrum examined in this work is obtained with the VLT/FORS2 spectrograph and published by \cite{Fynbo09}. The gas-phase abundances listed in Table~\ref{tab:metcol} are from \cite{Zafar19}. In addition to the parametric dust model we also obtain the best fit for the equivalent neutral hydrogen column density of $N_{\mathrm{H,X}} = (0.60\pm 0.05) \times 10^{22}$\,cm$^{-2}$. The SED fits well with a single power-law (with an F-test probability of 1.00) with a slope of $\beta_{\mathrm{opt,X}} = 0.62\pm 0.02$ and the resulting reduced $\chi^2$ together with the number of degrees of freedom is $\chi^2/(\mathrm{dof}) = 724/841$, with a null hypothesis probability of $100\%$ for the best fit model. \cite{Zafar11} have fitted the parametric dust model prescribed by \cite{Fitzpatrick07} to the same optical afterglow examined here. Our results are mostly consistent with their best fit model, except that we find a smaller visual extinction of $\approx 0.5$\,mag compared to their estimate of $A_V \approx 1.2$\,mag. This discrepancy is likely due to larger best fit value for $R_V$ of $\approx 3.3$ compared to the results by \cite{Zafar11} who found a slightly steeper slope for the extinction curve of $R_V\approx 2.9$.

\subsection*{GRB\,080607}

\begin{figure} %[!ht]
	\centering
	\epsfig{file=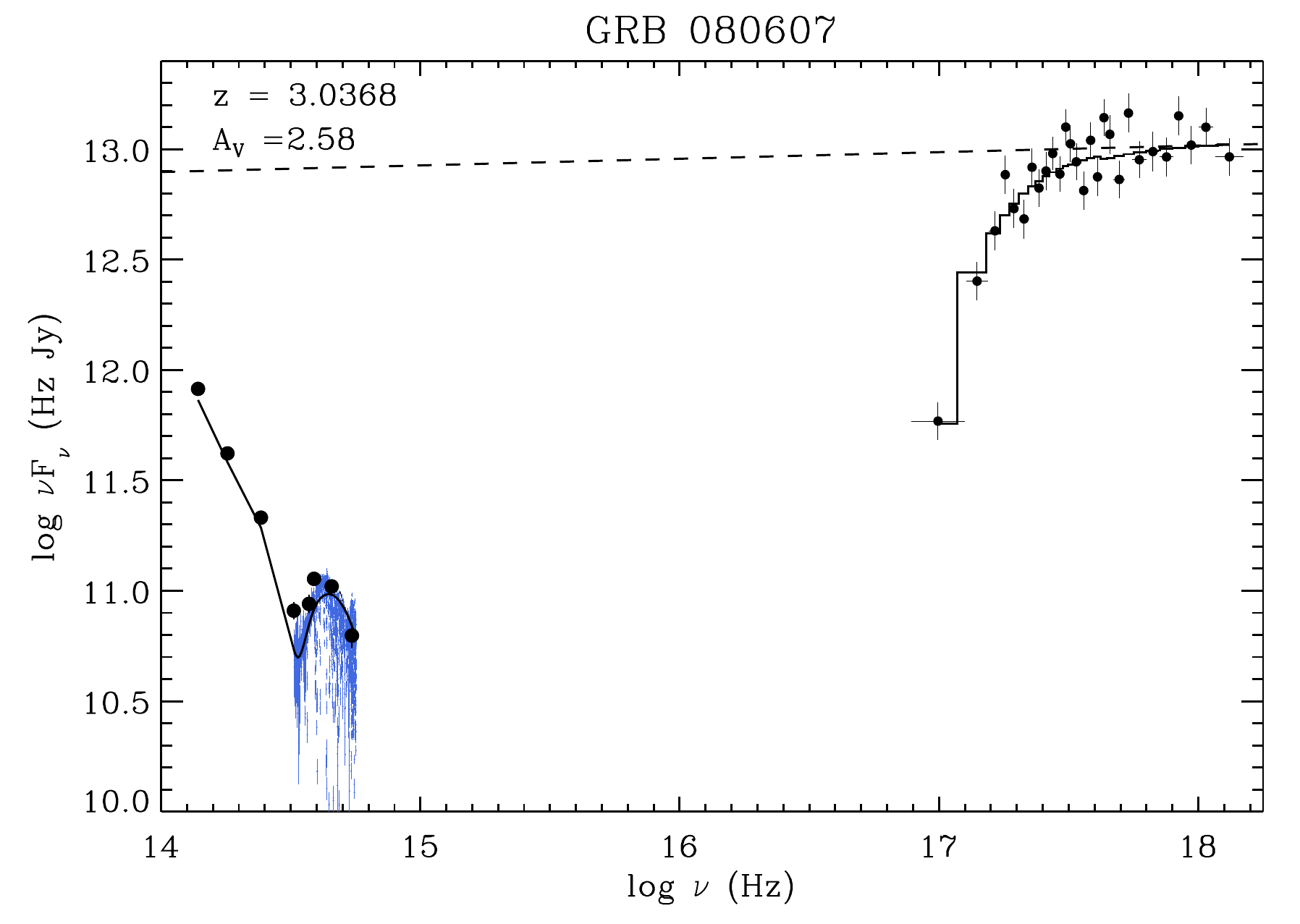,width=8.5cm}
	\caption{Same plot connotation as in Fig.~\ref{afig:061121} but for GRB\,080607 observed with the Keck/LRIS.}
	\label{afig:080607}
\end{figure}

This GRB was first detected by \textit{Swift} and reported by \cite{Mangano08}. For this case we derived the extinction curve parameters listed in Table~\ref{tab:ext} and the best fit model and $A_V$ are shown in Fig.~\ref{afig:080607}. The spectrum examined in this work is obtained with the Keck/LRIS spectrograph and published by \cite{Prochaska09,Fynbo09}. In addition to the parametric dust model we also obtain the best fit for the equivalent neutral hydrogen column density of $N_{\mathrm{H,X}} = (3.79^{+0.24}_{-0.22}) \times 10^{22}$\,cm$^{-2}$. The SED fits well with a single power-law (with an F-test probability of 1.00) with a slope of $\beta_{\mathrm{opt,X}} = 0.99\pm 0.05$ and the resulting reduced $\chi^2$ together with the number of degrees of freedom is $\chi^2/(\mathrm{dof}) = 326/481$, with a null hypothesis probability of $100\%$ for the best fit model. This GRB presents the first detection of H$_2$ in the optical afterglow spectrum, shows a clear sign of the 2175\,\AA~extinction bump \citep{Prochaska09} and is one of the most luminous bursts ever detected \citep{Perley11}. \cite{Zafar11} have fitted the parametric dust model prescribed by \cite{Fitzpatrick07} to the same optical afterglow examined here. Our results are consistent with their best fit model, but we will use the results from fitting the SED with the F\&M\,90 parametrization here for consistency.

\subsection*{GRB\,120119A}

This GRB was first detected by \textit{Swift} and reported by \cite{Beardmore12}. The data presented here were obtained with the VLT/X-shooter spectrograph, published by \cite{Selsing18} and have been examined by \cite{Japelj15} and \cite{Zafar18b}. We adopt the extinction curve parameters from \citet[][see Table~\ref{tab:ext}]{Zafar18b} since they fit for the full F\&M\,90 parametrization \citep[whereas][only fit for $A_V$ assuming either an MW, SMC or LMC-like extinction curve]{Japelj15}. While both found a significant visual extinction, \cite{Japelj15} claimed a detection of the 2175\,\AA~extinction bump, although the best-fit LMC extinction curve overpredicts the strength of the bump. Modelling the full F\&M\,90 parametric extinction law does not show a significant 2175\,\AA~bump, however \citep[see][]{Zafar18b}, and were are thus only able to provide upper limits on $A_{\mathrm{bump}}$. We adopt the column densities of H\,\textsc{i} and the low-ionization lines listed in Table~\ref{tab:metcol} from \cite{Wiseman17}.

\subsection*{GRB\,120815A}

This GRB was first detected by \textit{Swift} and reported by \cite{Pagani12a}. The data presented here were obtained with the VLT/X-shooter spectrograph and published and examined by \cite{Kruhler13} but we adopt the extinction curve parameters from \citet[][see Table~\ref{tab:ext}]{Zafar18b}. The GRB is found to exhibit a moderate amount of extinction with $A_V = 0.1 - 0.3$ mag and does not show any indication of the 2175\,\AA~extinction bump \citep{Kruhler13,Japelj15,Zafar18b}. We adopt the column densities of H\,\textsc{i} and the low-ionization lines listed in Table~\ref{tab:metcol} from \cite{Kruhler13}.

\subsection*{GRB\,121024A}

This GRB was first detected by \textit{Swift} and reported by \cite{Pagani12b}. The data presented here were obtained with the VLT/X-shooter spectrograph and published and examined by \citet{Friis15}. We adopt the extinction curve parameters from \citet[][see Table~\ref{tab:ext}]{Zafar18b} for this burst. \cite{Friis15} found a very steep extinction curve ($R_V > 15$) is required to explain the SED, but that is by assuming that the SED-inferred amount of dust, $A_V$, can be derived from the dust depletion and metallicity \citep[which is typically not the case;][]{DeCia13,Wiseman17}. Instead, \cite{Zafar18b} found that the GRB is well-fitted with a single power-law and a featureless extinction curve with $R_V \approx 2.8$. We adopt the column densities of H\,\textsc{i} and the low-ionization lines listed in Table~\ref{tab:metcol} from \cite{Friis15}.

\subsection*{GRB\,150403A}

\begin{figure}%[!ht]
	\centering
	\epsfig{file=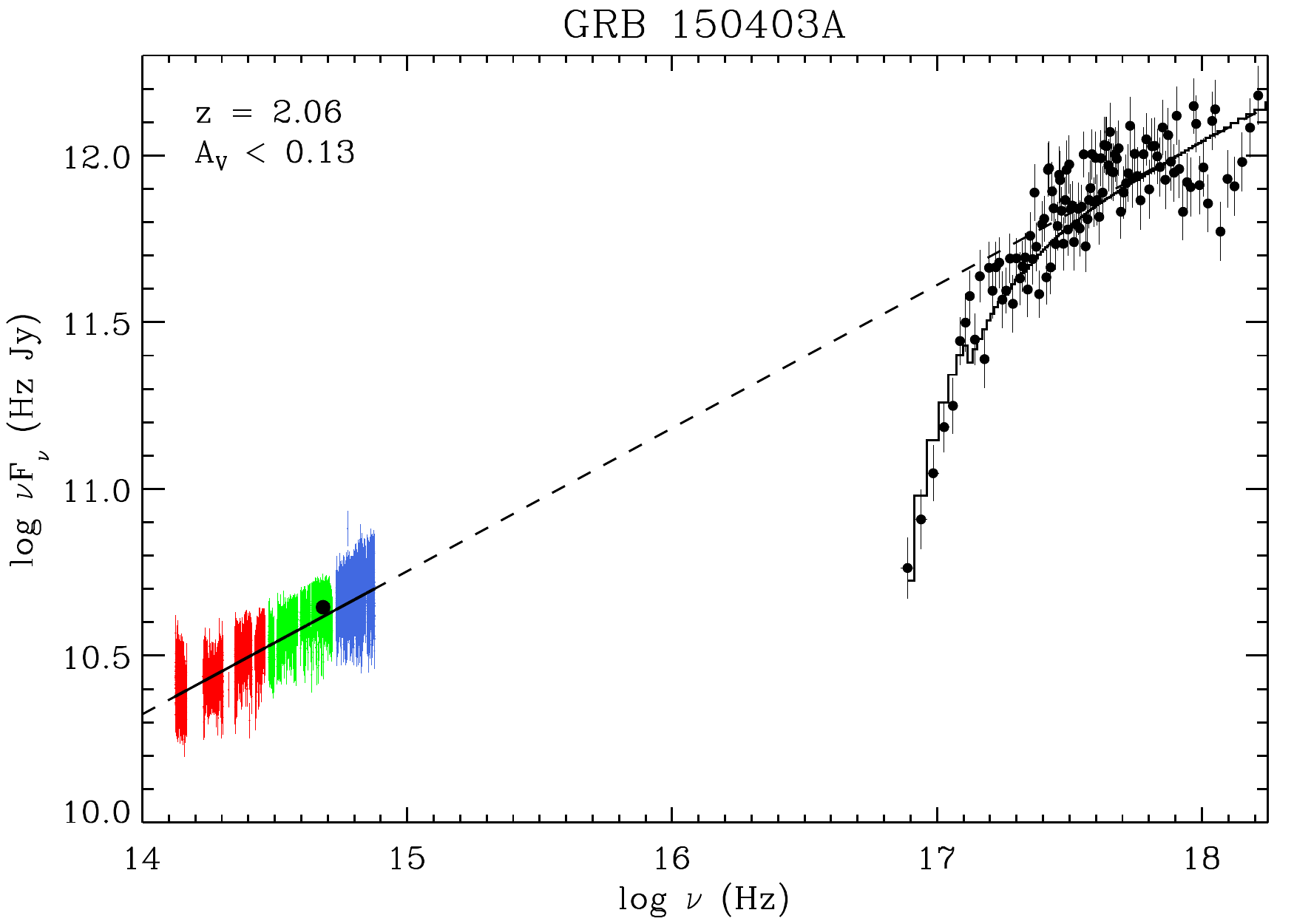,width=8.5cm}
	\caption{Same plot connotation as in Fig.~\ref{afig:061121} but for GRB\,150403A observed with the VLT/X-shooter. The blue, green and red colours correspond to the UVB, VIS and NIR arm spectra, respectively.}
	\label{afig:150403a}
\end{figure}

This GRB was first detected by \textit{Swift} and reported by \cite{Lien15}. The data presented here were obtained with the VLT/X-shooter spectrograph and published by \cite{Selsing18}. The gas-phase abundances listed in Table~\ref{tab:metcol} are from \cite{Bolmer18}. For this case we derived the extinction curve parameters listed in Table~\ref{tab:ext} and the best fit model and $A_V$ are shown in Fig.~\ref{afig:150403a}. In addition to the parametric dust model we also obtain the best fit for the equivalent neutral hydrogen column density of $N_{\mathrm{H,X}} = (0.22^{+0.12}_{-0.11})\times 10^{22}$\,cm$^{-2}$. The SED fits well with a single power-law (with an F-test probability of 1.00) with a slope $\beta_{\mathrm{opt,X}} = 0.57^{+0.14}_{-0.12}$ and the resulting reduced $\chi^2$ together with the number of degrees of freedom is $\chi^2/(\mathrm{dof}) = 31652/41269$, with a null hypothesis probability of $100\%$ for the best fit model.

\subsection*{GRB\,180325A}

This GRB was first detected by \textit{Swift} and reported by \cite{Troja18}. The data presented here were obtained with the VLT/X-shooter spectrograph and published and examined by \cite{Zafar18a}. We adopt the extinction curve parameters listed in Table~\ref{tab:ext} and the column densities of H\,\textsc{i} and the low-ionization lines listed in Table~\ref{tab:metcol} from their work. This GRB exhibit a clear presence of the 2175\,\AA~extinction bump, first noted in the low-resolution NOT/ALFOSC spectra by \cite{Heintz18a} obtained only 11 minutes post-burst. Subsequently, \cite{DAvanzo18} followed up this GRB with the VLT/X-shooter, confirming the detection of the 2175\,\AA~extinction bump and detected several absorption lines and the nebular emission lines [O\,\textsc{ii}], [O\,\textsc{iii}], and H$\alpha$. \cite{Zafar18a} found that the four different epochs of the GRB afterglow show consistent dust properties. We will use the best fit extinction curve parameters from the first epoch VLT/X-shooter spectrum in this work.

%%%%%%%%%%%%%%%%%%%%%%%%%%%%%%%%%%%%%%%%%%%%%%%%%%

% Don't change these lines
\bsp	% typesetting comment
\label{lastpage}
\end{document}